\documentclass{article}
\usepackage{spconf,amsmath,graphicx}

\DeclareMathOperator*{\argmin}{arg\,min}
\usepackage{cite}
\usepackage{graphicx}
\usepackage{amsmath,amssymb,amsfonts}
\usepackage{algorithmic}
\usepackage[ruled,vlined]{algorithm2e}
\usepackage{graphicx}
\usepackage{textcomp}
\usepackage{xcolor}
\usepackage{bbm}
\usepackage{multirow}
\usepackage{color,soul}
\usepackage[switch]{lineno}
\usepackage{adjustbox}
\usepackage{hyperref}
\newcommand*{\affmark}[1][*]{\textsuperscript{#1}}

\title{DynImp: Dynamic Imputation for Wearable Sensing Data Through Sensory and Temporal Relatedness}
\name{Zepeng Huo\affmark[1], Taowei Ji\affmark[1], Yifei Liang\affmark[1], Shuai Huang\affmark[2], Zhangyang Wang\affmark[3], Xiaoning Qian\affmark[1], Bobak Mortazavi\affmark[1]\thanks{Corresponding author: Zepeng Huo \{guangzhou92@tamu.edu\}. This project is in part supported by the Defense Advanced Research Projects Agency under grand FA8750-18-2-0027.}}
\address{\affmark[1]Texas A\&M University \\
\affmark[2]University of Washington \\
\affmark[3]University of Texas at Austin}
\begin{document}
%
\maketitle
\begin{abstract}
In wearable sensing applications, data is inevitable to be irregularly sampled or partially missing, which pose challenges for any downstream application. An unique aspect of wearable data is that it is time-series data and each channel can be correlated to another one, such as x, y, z axis of accelerometer. We argue that traditional methods have rarely made use of both times-series dynamics of the data as well as the relatedness of the features from different sensors. We propose a model, termed as DynImp, to handle different time point's missingness with nearest neighbors along feature axis and then feeding the data into a LSTM-based denoising autoencoder which can reconstruct missingness along the time axis. We experiment the model on the extreme missingness scenario ($>50\%$ missing rate) which has not been widely tested in wearable data. Our experiments on activity recognition show that the method can exploit the multi-modality features from related sensors and also learn from history time-series dynamics to reconstruct the data under extreme missingness.
\end{abstract}
\begin{keywords}
Wearable Sensing, Imputation, Neural Networks
\end{keywords}
\vspace{-2mm}
\section{Introduction}
\label{sec:intro}

The ubiquity of wearable sensor data allows for daily well-being tracking by recognizing the user activities \cite{huo2020uncertainty}, but models trained to enable this tracking often falter in performance when deployed in real-world environments, due to data quality and consistency issues (including missing data), sensor noise, and/or lack of user adherence poses \cite{solis2019human, kim2013faulty}. While adherence is a potential behavioral issue, sensor noise (which, if identified, can be eliminated and treated as missing) and missing data are often viewed as a key step in enabling real-world data tracking, producing a broad range of imputation techniques for missing data \cite{faizin2019review}. For activity recognition application, due to the motion artifact or other reasons, missing data can be a particularly prominent issue and dampen the performance of machine learning models \cite{michael2017activity}.

\vspace{-4mm}
Traditional missing value imputation techniques include filling with the mean value of a feature, or using forward or backward imputation~\cite{lipton2016modeling}. These static imputation methods do not reflect the underlying time-varying dynamics of data and therefore may bias downstream predictions based on the estimated likelihood of imputed values. Additionally, in wearable sensing applications, the different sensing channels are often capturing the same events of interest, so multiple imputation approaches may be suited to addressing this kind of missing data \cite{azur2011multiple, lin2020filling}. However, these techniques have remained static in imputation as well. For example, a Multi-layer Perceptron (MLP)-based model for irregular time-series data handling that accounts for such multiple imputation, still interpolates only a single channel \cite{abedin2019sparsesense}, handling time-dynamics but not able to model across all channels (in other words, multiple models would be needed). On the other hand, MissForest, a tree-based machine learning method that predicts missing values from related, non-missing data \cite{stekhoven2012missforest}, overcomes these limitations but does not make use of time-varying dynamics. In addition, the rate of missing data in traditional studies is relatively low, ranging from 1.98\% to 50.65\% \cite{madhu2019novel, lipton2016modeling, stekhoven2012missforest}. A stronger imputation technique is needed for real-world applications with severe missingness testing.

\begin{figure*}[t!]
\centering
\includegraphics[width=0.85\textwidth]{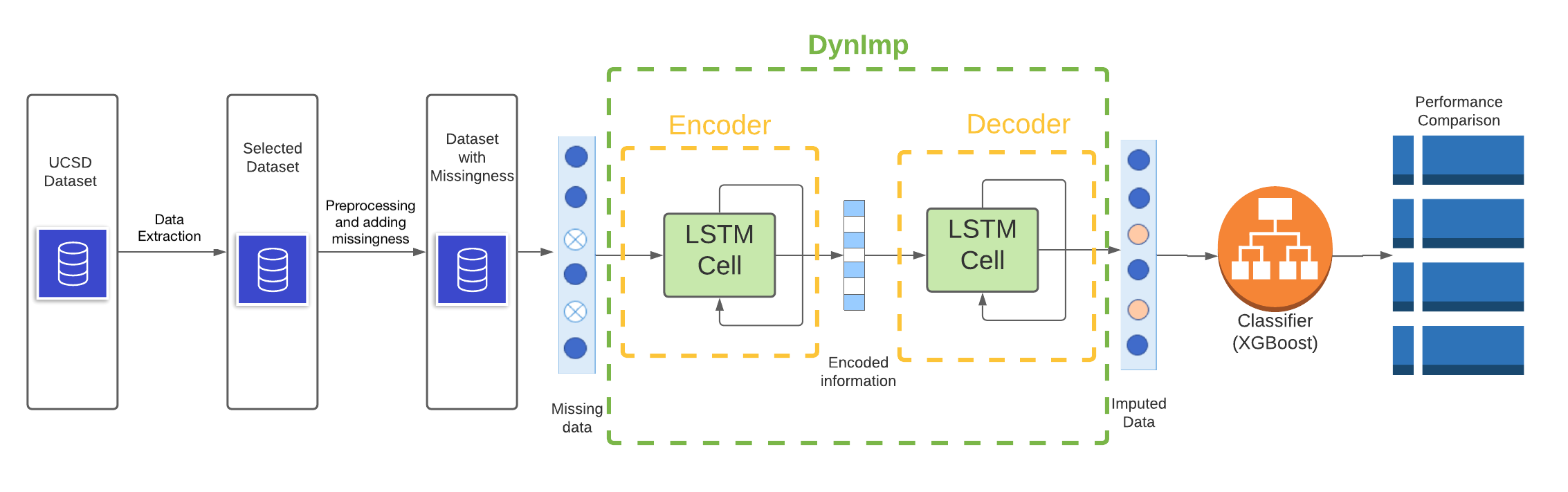}
\vspace{-3mm}
\caption{Pipeline for Dynamic Imputation}
\label{fig:pipline}
\end{figure*}

\vspace{-7mm}
We propose to use a long-short-term-memory-based denoising autoencoder (LSTM-DAE) to learn more robust imputation strategies for remote sensing data. The model has an encoder and decoder architecture to embed signal data, then this encoded information is fed into the LSTM network to learn the time-varying dynamics of the data. The overall structure is shown in Fig. \ref{fig:pipline}. This architecture robustly imputes missing data from related channels and latest dynamics being measured by all sensor channels, even in the presence of high rates of missing data. 
In order to demonstrate the utility of using both time-series dynamics and feature relatedness, we experiment on datasets with both inherent missing values and increased missing data (up to 60\% missing across all the channels on a dataset with inherent 66\% missingness), which surpasses traditional missing rate study by a large margin.


\section{Related Work}
\vspace{-4mm}
There has been numerous studies on missing-data approaches which brought about promising results for down-stream modeling. MissForest can handle mixed types of features and of missing data value by using a tree-based method \cite{stekhoven2012missforest}. In \cite{de2013predicting}, the authors proposed a bi-clustering based data imputation technique using the mean squared residual metric that estimates the degree of coherence between each recorded cell of the dataset. In \cite{yan2015missing}, the authors present the an imputation method for missing data value in Internet of Things (IOT) device data, by applying context-based linear mean, binary search method as well as a Gaussian mixture model. There are many deep learning based methods as well, in \cite{abedin2019sparsesense}, the authors a multi-layer perceptron coupled with interpolation technique. In \cite{lipton2016modeling}, the authors chose not to directly impute, but rather treated the missing values as extra source of information and used an LSTM model to train and predict the time series data. This work will evaluate the strengths of each of these techniques in comparison to the work proposed here.
\vspace{-2mm}
\section{Methodology}
\vspace{-4mm}
Our proposed architecture is illustrated in Fig. 1. We will discuss the building components of them in each following sections.
The architecture is split into an encoder/decoder architecture. The time-series information is modeled by a LSTM-based autoencoder, coupled with KNN-padding to exploit the feature relatedness in wearable data.

\vspace{-5mm}
\subsection{Denoising Autoencoder}
\vspace{-1mm}
The aims for autoencoder are delivered by learning representations (encoding) of a set of data and reconstructing (decoding) the data from these representations. 
Through this encoding and decoding process, the network can take care of data missingness and possible signal noise that lies within the data. A generic autoencooder would be formulated as:
\begin{equation}
\begin{aligned}
    \theta^*, \theta'^* &= \argmin_{\theta, \theta'} \frac{1}{n}\sum_{i=1}^n L(\mathbf{x}^{(i)}, \mathbf{z}^{(i)}) \\
    &=\argmin_{\theta, \theta'} \frac{1}{n}\sum_{i=1}^n L(\mathbf{x}^{(i)}, g_{\theta'}(f_\theta(\mathbf{x}^{(i)}))),
\end{aligned}    
\end{equation}
where the encoding function is $\mathbf{y}=f_\theta(\mathbf{x})=f(\mathbf{Wx}+\mathbf{b})$, and the decoding function to map the latent representation back is $\mathbf{z}=g_{\theta'}(\mathbf{y})=g(\mathbf{W'y}+\mathbf{b})$. $L$ is represented as the loss function, which can be taken as the form of reconstruction cross-entropy: \vspace{-3mm}
\begin{equation}
L(\mathbf{x}, \mathbf{z}) = - \sum_{k=1}^d [\mathbf{x}_k\log \mathbf{z}_k + (1-\mathbf{x}_k)\log(1-\mathbf{z}_k)].
\end{equation}
The denoising autoencoder is to enforce some extra noise on the input vector so the input vector will be \textit{corrupted}. 
This is based on the assumption that the core data representation from a large amount of training example will stay relatively stable even with background noises so the decoder will learn to distinguish that from the stochastic noise \cite{vincent2008extracting, huo2019sparse}. 
The stochastic mapping can be written as $\tilde{\mathbf{x}} \sim q_D(\tilde{\mathbf{x}}|\mathbf{x})$. The mapping function can take various forms to corrupt the input vector. We used a stochastic dropout as the function 
\begin{equation}
\begin{aligned}
    & r_j \sim \text{Bernoulli}(p), \\
    & \tilde{\mathbf{x}} = \mathbf{r} \cdot \mathbf{x},
\end{aligned}
\end{equation}
where $\mathbf{r}$ is a vector of independent
Bernoulli random variables, each of which has probability $p$ of being 1. And $*$ is an element-wise product, so this vector is sampled and multiplied element-wise with the input to give the \textit{corrupted} signal. 

\vspace{-3mm}
\subsection{Padding for Denoising Autoencoder}
\vspace{-1mm}
Traditional methods for training the neural networks would usually pad missing data with interpolated values or simply zeros \cite{swietojanski2014convolutional}. However for a dynamic imputation scenario, the missingness should be captured with an adaptive manner, which means as time goes on, the imputation should also adjust with newer data coming in.
We propose to integrate K-NN imputation with the neural network. Therefore, in each time point, the model will have different nearest neighbors, and the combination with autoencder can be used to capture this higher level of dynamics. Therefore our hidden layer function can be further written as: 
\begin{equation}
    h(\tilde{\mathbf{x}}^{(i)}) = g[\mathbf{W'}\cdot ((\mathbf{M}^{t}\odot \tilde{\mathbf{x}}^{(i)})+\mathbf{P})+\mathbf{b}],
\end{equation}
where $\mathbf{M}^t$ is a masking matrix for indicating the missingness in input data, $\odot$ is element-wise production operator. Assume we have $i$-th data entry missing on $t$-th time point: 
\begin{equation}
    \mathbf{M}^t = \mathbbm{1}_{(m, l)} \left\{
                \begin{array}{lr}
                  0 \ \ \ \ \   \text{if } i=m, t=l, \\
                  1 \ \ \ \ \  \text{otherwise,}
                \end{array}
              \right.
\end{equation}
where $\mathbbm{1}$ is an indicator function, $m$ and $l$ are the indicator indices of the masking matrix, representing the $x$ and $y$ axes respectively. Furthermore the imputation matrix with nearest neighbor padding $\mathbf{P}$ is as follows:
\begin{equation}
    \mathbf{P}^t = \left\{
                \begin{array}{lr}
                  \frac{\sum_j \mathbf{x}^{(j)}}{k} \ \ \ \ \   \text{if $j$ in top $k$ neighbors of $i$}  , \\
                  0 \ \ \ \ \  \text{otherwise.}
                \end{array}
              \right.
\end{equation}
where $i$ is the index of interest (i.e. missing) and $j$ is a set of indices representing the top $k$ neighbors of $i$. The missing value would be from the mean of $k$ closes neighbor's average, where the distance is measured in Euclidean distance. Here we have constructed our denoising autoencoder with nearest neighbor padding technique for time-series data.



\vspace{-3mm}
\subsection{LSTM-based Autoencoder}
\vspace{-1mm}
In order to exploit time-series dynamics, we propose that the fully connected layer in traditional autoencoder would need to replace the function $\mathbf{y}=f_\theta(\mathbf{x})=f(\mathbf{Wx}+\mathbf{b})$ to a LSTM cell, where each input feature $\mathbf{x}$ has a time-stamp, i.e. $\mathbf{x}_t$.

First the LSTM cell will generate a decision vector and select the candidate information.
For the current time stamp $t$, the vector $I_t$ is a function of last hidden state $h_{t-1}$ and input feature $x_t$, and the output gate $f_o$ will generate the hidden state $h_t$ conditioned on the output:
\begin{equation}
\begin{aligned}
    I_t &= f_i (w_ix_t+w_ih_{t-1}+b_i), \\
    F_t &= f_g(w_gx_t + w_gh_{t-1}+b_g), \\
    \tilde{C}_t &= f_C(w_cx_t + w_ch_{t-1} + b_c), \\
    C_t &= C_{t-1}F_t + \Tilde{C}_tI_t, \\
    Y_t &= f_o(w_ox_t + w_oh_{t-1} + b_o), \\
    h_t &= Y_tf_h(C_t).
\end{aligned}
\end{equation}
This hidden state will replace the original encoding function output $f_\theta(\mathbf{x})=f(\mathbf{Wx}+\mathbf{b})$ and thus will become the encoded information that is later fed into decoder for reconstruction, following similar steps as above for time-series imputation.


\section{Experiment}

\begin{table*}[t!]

\centering
\caption{Results for comparison between baseline imputation and dynamic imputation (in BA and its  confidence interval)}

\label{tab:result}
\begin{adjustbox}{width=0.9\textwidth}
\begin{tabular}{|l|l|l|l|l|l|l||l|}
\hline

\multicolumn{2}{|c|}{Level of missingness}  & \multicolumn{1}{c|}{Filled Mean} & \multicolumn{1}{c|}{kNN imputer} & \multicolumn{1}{c|}{Missforest} & \multicolumn{1}{c|}{SparseSense [MLP]} & \multicolumn{1}{c||}{Indicator Variable [LSTM]}  & \multicolumn{1}{c|}{DynImp [LSTM-DAE]} \\ \hline
\multirow{2}{*}{mild}&
10\%    & 0.8259 $\pm$ 0.0069     & 0.8288 $\pm$ 0.0046   & \textbf{0.8458 $\pm$ 0.0027}  & 0.8213 $\pm$ 0.0032    & 0.8248 $\pm$ 0.0085         & 0.838 $\pm$ 0.0095   \\

& 20\%    & 0.8065 $\pm$ 0.0051     & 0.8001 $\pm$ 0.0069   & 0.8364 $\pm$ 0.0052 & 0.8108 $\pm$ 0.0077  & 0.8081 $\pm$ 0.0078         &  \textbf{0.8390 $\pm$ 0.0045}  \\
\hline
\multirow{2}{*}{medium}&
30\%    & 0.7871 $\pm$ 0.0043     & 0.7802 $\pm$ 0.00097  & 0.8131 $\pm$ 0.0046  & 0.7901 $\pm$ 0.0060& 0.7852 $\pm$ 0.0060        & \textbf{0.8318 $\pm$ 0.0012}  \\

& 40\%    & 0.7627 $\pm$ 0.0092    & 0.7570 $\pm$  0.0106  & 0.7918 $\pm$ 0.0072& 0.7647 $\pm$ 0.0079   & 0.7624 $\pm$ 0.0105         & \textbf{0.826 $\pm$ 0.0042}    \\
\hline
\multirow{2}{*}{severe}&
50\%    & 0.7367 $\pm$ 0.0087    & 0.7351 $\pm$ 0.0067   & 0.7541 $\pm$ 0.0066& 0.7401 $\pm$ 0.0095   & 0.7401 $\pm$ 0.0104        & \textbf{0.831 $\pm$ 0.0071}  \\

& 60\%    & 0.7043 $\pm$ 0.0085   & 0.7087 $\pm$ 0.0086   & 0.7131 $\pm$ 0.0093  & 0.6907 $\pm$ 0.0122  & 0.6907 $\pm$ 0.0071        & \textbf{0.8304 $\pm$ 0.0070}\\
\hline
\end{tabular}

\end{adjustbox}
\end{table*}


\begin{table}[t!]

\centering
\caption{Results for variants of DynImp for missingness}

\label{ablation}
 \begin{adjustbox}{width=0.5\textwidth}
\begin{tabular}{|r||l|l|l|l|l|l|}
\hline                                                                                                                                                                                                       
DynImp Variations &  10\% & 20\% & 30\% & 40\% & 50\% & 60\% \\ \hline
 (0-padding) &      0.8262       & 0.8219   & 0.8239   &  0.8236 & 0.8288 & 0.8299       \\
 (Filled mean) &    0.8266      & 0.8224   & 0.8219   &  0.8245 & 0.8232 & 0.8254      \\
 (Interpolation) &  0.8284      & 0.8103   & 0.8291   &  0.8201 & 0.8263 & 0.8290        \\
 (kNN) &            \textbf{0.8380}      & \textbf{0.8390}   & \textbf{0.8318}   &  \textbf{0.8260}  &  \textbf{0.8310} & \textbf{0.8304}        \\
\hline
\end{tabular}

\end{adjustbox}
\end{table}

\vspace{-3mm}
\subsection{Dataset}
We use the UCSD ExtraSensory dataset for these experiments \cite{vaizman2018context}.
The UCSD ExtraSensory dataset contains data from 60 users (34 female and 26 male). Data were collected through a user’s personal smart phone (34 iPhone, 26 android). The sensors include high-frequency motion reactive sensor (accelerometer, gyroscope, magnetometer, etc.). We chose this dataset because the it collects motion-reactive sensor data and the dynamics of one sensor has potentiality to be recovered by another one due to underlying relatedness \cite{vaizman2018context}.
The inherent missingness is 66\% for each sensor per their description. we observe that this dataset has used a data collection app that \textit{performs a 20-second ``recording session'' automatically every minute}. The intuition was not given. We hypothesize that it could be from the battery consumption consideration or communication overhead in the smart phone. But in any case the data are presented with this blank period, considered as ``missingness'', which might not suffice as a `random missingness'. We have further randomly perturbed the data with varied missingness levels from 10\% to 60\% to evaluate different imputation strategies.
\vspace{-5mm}
\subsection{Experimental Setup}
\vspace{-1mm}
We selected a group of features that represent data collected widely from wearable sensors \cite{vaizman2018context}. We considered raw measurements for the accelerometer, gyroscope, and magnetometer across all three channels (axis). We also selected two features for the location sensor, which were the mean absolute longitude and latitude. The total number of features (including all the axes of each feature) is 14. We use a sliding window technique, where the window length is 24 minutes and features are grouped by 1 minute intervals. So for each window we will use the features from each sensor to predict the body movement label. 
Regarding to the labels, we have two concurrent labels for the users which are 4 body movements and 4 phone locations (where on the body).
The 4 body movements labels which are mutually exclusive in nature and combined into one multiple class label, which are 'LYING\_DOWN', 'SITTING', 'FIX\_walking', and 'others'. With these 4 body movement labels, as well as their combinations with the 4 phone locations on the body, we have 16 labels in all the combinations.
We used classification report from sklearn, and recorded balanced accuracy (BA) as the  model performances evaluation \cite{vaizman2018context}. 
In order to achieve a stabilized result and perform fair comparisons, we composed each kinds of testing with different setup mentioned above 10 times (each time with different seed for random missingness generation) and average the BA in validation set, and the corresponding confidence interval.
Regards to different baseline imputation method and different state-of-the-arts, we picked:
\vspace{-7mm}
\begin{itemize}
    \setlength\itemsep{-0.5em}
    \item Filled mean: each missing value is computed from the average of the time-series for that sensor feature.
    \item kNN imputer: each missing value is computed by finding the nearest neighbors. The distance is measured by euclidean distance of the remaining values for that sensor feature to the neighbors.
    \item Missforest: a tree-based imputation method that learns to predict the missing values
    \item SparseSense (MLP) \cite{abedin2019sparsesense}: a neural network based framework that combines interpolation and Multi-Layer Perceptron to predict the missing value
    \item Indicator Variable (LSTM) \cite{lipton2016modeling}: a LSTM-based method to use missing value as extra features, an indicator (masking) feature vector to indicate the imputed value is missing or not, where 1 is missing and 0 is not. 
\end{itemize}
\vspace{-1.5mm}

\vspace{-3mm}

\subsection{Results}
We put the results of experiment into Table \ref{tab:result}, illustrating the comparison between baselines. The models performance were introduced regarding the balanced accuracy (BA) \cite{vaizman2018context} in validation set. The row denotes for increasing random missingness generated in the data.  
From the result tables overall we can discern a pattern for traditional imputation methods: as the level of missingness increases, the model performance decreases drastically. We notice that Missforest can perform relatively well albeit marginal, when the missingness is mild, but still degrades quickly when the missingness enlarges. We hypothesize that when coupled with similar tree-based classifier, i.e. XGBoost, the Missforest can reconstruct the data that is best for a boosting classifier. One previous study  \cite{madhu2019novel} showed the Missforest coupled with XGboost can perform relatively well on mild missingness scenario (2.19\% to 13.63\%). But for our model, it outperforms not only the traditional methods but also the methods that utilizes the neural networks such as MLP and LSTM. The performs can hold even with extremely severe missingness.


\begin{figure}[h]
\centering
\includegraphics[width=0.41\textwidth]{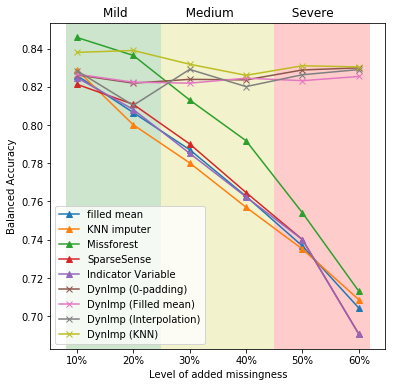}
\vspace*{-5mm}\caption{Results comparison}
\label{fig:results}
\end{figure}


\vspace{-3mm}
\subsection{Ablation Study}

We have established the improvement of our model by making use of LSTM-DAE pipeline. Next in order to study the necessity of using nearest neighbor as the padding strategy in our architecture we conduct a comparison among the variants: \vspace{-5mm}
\begin{itemize}
\setlength\itemsep{-0.1cm}
    \item DynImp (0-padding): the missingness is replaced by zeros of the time-series for that sensor feature before feeding to DynImp for training \cite{swietojanski2014convolutional}
    \item DynImp (Filled mean): the missingness is replaced by average of the time-series for that sensor feature before feeding to DynImp for training \cite{lipton2015learning}
    \item DynImp (Interpolation): the missingness is replaced by interpolation of before and after values before feeding to DynImp for training \cite{abedin2019sparsesense}
    \item DynImp (kNN): the missingness is replaced by kNN-imputer for $k=5$ neigbhors, before feeding to DynImp
\end{itemize}\vspace{-3mm}
As we can see in Table \ref{ablation} the superior performance of our proposed scenario, even with the same network architecture, the nearest neighbor method can further push the LSTM-DAE to have better performance than baselines.
A clear illustration of all methods is also shown in Fig \ref{fig:results}. 

\vspace{-3mm}
\section{Conclusion}

In this paper we proposed a dynamic imputation technique for remote sensing data. The method is through a trainable mechanism by the use of deep learning model to learn the missing dynamics along the time axis to impute the missing data. The model shows strong performance compared to baselines. While not particularly outstanding in mild missingness scenario, the model can maintain high prediction accuracy in some extreme missingness scenarios. We also tested on variations of the model to find out the best performing one. This dynamic imputation can be classifier-agnostic so it can be compatible with any downstream methods. In the future we aim to study how does the model perform in other scenarios in terms of different data types and features.

\bibliographystyle{IEEEbib}
\bibliography{bib}

\end{document}